\documentstyle[psfig,aps]{revtex}

\catcode`\@=11
\def\marginnote#1{}

\def\ifmath#1{\relax\ifmmode #1\else $#1$\fi}

\def\bold#1{\setbox0=\hbox{$#1$}%
     \kern-.025em\copy0\kern-\wd0
     \kern.05em\copy0\kern-\wd0
     \kern-.025em\raise.0433em\box0 } 

\def\GENITEM#1;#2{\par\vskip6pt \hangafter=0 \hangindent=#1
   \Textindent{$ #2$ }\ignorespaces}

\newcount\hour
\newcount\minute
\newtoks\amorpm
\hour=\time\divide\hour by60
\minute=\time{\multiply\hour by60 \global\advance\minute by-
\hour}
\edef\standardtime{{\ifnum\hour<12 \global\amorpm={am}%
    \else\global\amorpm={pm}\advance\hour by-12 \fi
    \ifnum\hour=0 \hour=12 \fi
    \number\hour:\ifnum\minute<100\fi\number\minute\the\amorpm}}
\edef\militarytime{\number\hour:\ifnum\minute<100\fi\number\minute}
\def\draftlabel#1{{\@bsphack\if@filesw {\let\thepage\relax
  \xdef\@gtempa{\write\@auxout{\string
    \newlabel{#1}{{\@currentlabel}{\thepage}}}}}\@gtempa
    \if@nobreak \ifvmode\nobreak\fi\fi\fi\@esphack}
     \gdef\@eqnlabel{#1}}
\def\@eqnlabel{}
\def\@vacuum{}
\def\draftmarginnote#1{\marginpar{\raggedright\scriptsize\tt#1}}
\def\draft{\oddsidemargin -.5truein
        \def\@oddfoot{\sl preliminary draft \hfil
        \rm\thepage\hfil\sl\today\quad\militarytime}
        \let\@evenfoot\@oddfoot \overfullrule 3pt
        \let\label=\draftlabel
        \let\marginnote=\draftmarginnote

\def\@eqnnum{(\theequation)\rlap{\kern\marginparsep\tt\@eqnlabel}%
\global\let\@eqnlabel\@vacuum}  }
\def\preprint{\twocolumn\sloppy\flushbottom\parindent 1em
        \leftmargini 2em\leftmarginv .5em\leftmarginvi .5em
        \oddsidemargin -.5in    \evensidemargin -.5in
        \let\@evenfoot\@oddfoot \overfullrule 3pt
        \let\label=\draftlabel
        \let\marginnote=\draftmarginnote

\def\@eqnnum{(\theequation)\rlap{\kern\marginparsep\tt\@eqnlabel}%
\global\let\@eqnlabel\@vacuum}  }
\def\preprint{\twocolumn\sloppy\flushbottom\parindent 1em
        \leftmargini 2em\leftmarginv .5em\leftmarginvi .5em
        \oddsidemargin -.5in    \evensidemargin -.5in
        \columnsep 15mm \footheight 0pt
        \textwidth 250mmin      \topmargin  -.4in
        \headheight 12pt \topskip .4in
        \textheight 175mm
        \footskip 0pt

\def\@oddhead{\thepage\hfil\addtocounter{page}{1}\thepage}
        \let\@evenhead\@oddhead \def\@oddfoot{} \def\@evenfoot{}
}
\def\titlepage{\@restonecolfalse\if@twocolumn\@restonecoltrue\onecolumn
     \else \newpage \fi \thispagestyle{empty}\c@page\z@
        \def\thefootnote{\fnsymbol{footnote}} }
\def\endtitlepage{\if@restonecol\twocolumn \else  \fi
        \def\thefootnote{\arabic{footnote}}
        \setcounter{footnote}{0}}  
\catcode`@=12
\relax

\def\be{\begin{equation}}
\def\ee{\end{equation}}
\def\br{\begin{eqnarray}}
\def\er{\end{eqnarray}}

\def\NPB#1#2#3{{\it Nucl.~Phys.} {\bf{B#1}} (19#2) #3}
\def\PLB#1#2#3{{\it Phys.~Lett.} {\bf{B#1}} (19#2) #3}
\def\PRD#1#2#3{{\it Phys.~Rev.} {\bf{D#1}} (19#2) #3}

\def\AP#1#2#3{{\it Ann.~Phys.} {\bf#1} (19#2) #3}

\def\PRD#1#2#3{{\it Phys.~Rev.} {\bf{D#1}} (19#2) #3}

\def\AP#1#2#3{{\it Ann.~Phys.} {\bf#1} (19#2) #3}

\begin{document}

\topmargin-1.5cm  

\begin{titlepage}
\begin{flushright}
SUSX-TH/97-024\\ 
IEM-FT-169/97\\
\end{flushright}
\vspace{.2in}
\begin{center}

{\large{\bf On non-perturbative corrections to the K\"ahler potential}
}
\bigskip \\
{\large T.~Barreiro${}^a$\footnote{Work supported by JNICT (Portugal).}, 
B.~de Carlos${}^b$\footnote{Work supported by PPARC.} and 
E.~J.~Copeland${}^{c \dagger}$\\ 
\vskip 0.2in
{\it 
Centre for Theoretical Physics, University of Sussex, \\ Falmer, 
Brighton BN1 9QH, UK. \\}
}
\vskip 0.2in
{\it 
${}^a$ Email: {\tt mppg6@pcss.maps.susx.ac.uk} \\
${}^b$ Email: {\tt B.De-Carlos@sussex.ac.uk} \\
${}^c$ Email: {\tt E.J.Copeland@sussex.ac.uk} \\}
\vspace{.5in}
{\bf Abstract} \smallskip \end{center} \setcounter{page}{0}
We present the results of a detailed investigation into the 
consequences of adding specific string motivated non-perturbative 
corrections to the usual tree level K\"ahler potential in dilaton 
dominated scenarios. The success of the model is judged through our 
ability to obtain a realistic VEV for the dilaton $<{\rm Re} S> \sim 
2$, corresponding to the true minima of the scalar potential and being
associated with a reasonable value for the SUSY breaking scale via the
gravitino mass. The status of the so-called moduli problem is also
reviewed in each of the ansatze studied. Those include previous
proposals made in the context of both the chiral and the linear
multiplet formalisms to describe gaugino condensation, and a new ansatz
which shows explicitly the equivalence between the two.

\end{titlepage}

\newpage
 
\section{Introduction}
There are a number of pressing issues that any string based scenario
of supersymmetry (SUSY) breaking needs to address. They include the 
stabilisation of the dilaton with an acceptable phenomenological 
value as it has Planck mass suppressed couplings with all the matter 
fields, the stabilisation of the moduli fields fields as they 
determine the size of the extra dimensions, the succesful breaking of 
supersymmetry at low energies and the notoriously difficult 
cosmological constant problem. In many ways, these issues are related 
to each other. If the dilaton field is stabilised with anything other 
than zero potential energy associated with it then there exists a 
residual cosmological constant. Over the past few years as interest in
string phenomenology has exploded there has been much progress made in
addressing these issues, in particular in the context of gaugino
condensation as the source of SUSY breaking \cite{deren85},
but it is fair to say, no totally satisfactory 
resolution has yet been obtained. However a few steps forward
have been made and it is possible to describe hierarchical SUSY breaking
and a non-trivial potential $V(S)$ for the dilaton $S$ (recall
Re $S \sim g^{-2}_{\rm gut}$) \cite{krasni87}. With two hidden 
condensates and the presence of hidden matter it was possible for
the dilaton to dynamically acquire a reasonable vaccum expectation 
value (VEV) (Re $S \sim 2$), with SUSY being broken at the correct
scale $m_{3/2} \sim 1 {\rm TeV}$, where $m_{3/2}$ is the gravitino mass. 
Unfortunately the dilaton was always stabilised leaving a negative 
cosmological constant \cite{decar93}. Moreover, the potential barrier 
which defined the minima was 
extremely small on the weak coupling side of the field, leading
to potential problems associated with stabilising the field in the
context of cosmology \cite{brust93}. 

More recently the issue of stabilising the dilaton an its relation to 
gaugino condensation has been considered
in the light of alternative approaches \cite{others};
in particular, Casas \cite{casas96} has investigated possible string 
induced non-perturbative corrections that the K\"ahler potential is likely to 
experience. Keeping the K\"ahler potential arbitrary he obtained 
generalised expressions for the soft breaking terms. Specific forms for the 
corrections were later introduced, motivated from string considerations. 
An important feature that emerged was that it 
became possible to obtain a minimum of the potential at the realistic value 
$S \sim 2$ with just one condensate, although it did not correspond to 
$V(S)=0$. 

In this paper, we examine in far greater detail than has previously 
been undertaken the predictions arising out of these non-perturbative 
corrections. In section 2  we introduce the analysis of the one 
condensate model and describe the types of non-perturbative corrections 
introduced by Casas \cite{casas96}. We then describe the various constraints 
imposed on the model, hence on the available parameters by demanding 
physically sensible values for $<S_0>$, $m_{3/2}$ and $V(S_0)=0$ at
the minimum. It turns out to be non trivial obtaining all three 
conditions together, and in the cases where it can be done, the price that 
is paid is a very low potential barrier for the dilaton to overcome
and large values of the parameters. Demanding the dilaton is 
properly stabilised at a reasonable value generally causes the 
gravitino mass to become
unrealistically large. However, the moduli problem could be solved by this 
mechanism. The dilaton mass turns out 
to be inversely proportional to $K''$, the second derivative of the
K\"ahler potential, which approaches zero at the 
minimum of the potential. Hence the mass becomes large compared to 
the gravitino mass. In section 3 we relate our results with those obtained by 
Bin\'etruy, Gaillard and Wu \cite{binet97a,binet97b} who also investigate 
non-perturbative corrections but based on the linear multiplet 
formalism. The comparisons are very encouraging, they demonstrate the
equivalence of the two approaches in that it is possible to relate the
potentials in the two cases. In particular we show that the results in
\cite{binet97a,binet97b} correspond to those in \cite{casas96} over 
a particular range of parameters. We conclude in section 4.

\section{The chiral multiplet formalism}

The scalar potential in any N=1 SUGRA model is given by:
\be
V = e^K |W|^2 \left[ \left(K^i + \frac{W^i}{W} \right) (K_i^j)^{-1}
\left( K_j + \frac{\bar{W}_j}{\bar{W}} \right) - 3 \right]
\label{pot}
\ee
where $K$ is the K\"ahler potential, $W$ is the superpotential and the
subindices $i$, $j$ represent derivatives of these two functions with 
respect to the different fields.

We are going to concentrate on models of SUSY breaking via gaugino 
condensation \cite{deren85}. These have been extensively studied in 
the literature, and we know that the form of the superpotential for
one condensate must be:
\be
W(S,T) = C e^{-\alpha S}/ \eta^6(T)
\label{sup}
\ee
where $\alpha = 8 \pi^2/N$, $C = -N/(32 \pi^2 e)$ if the
hidden sector group is SU(N) with no matter fields. The 
$\eta^{-6}(T)$ dependence ensures the correct transformation of $W$ 
under the SL(2,Z) target space modular symmetry,
and, in particular, we shall concentrate in dilaton-dominated
scenarios \cite{kaplu93}, as we are mainly concerned with the $S$
dependence of the scalar potential. Therefore in eq.~(\ref{pot})
we shall assume that $i=S$,$j= \bar{S}$, and then the scalar potential
reduces to:
\be
V = e^K  \left| \frac{C e^{- \alpha S} }{\eta^6(T)} \right|^2 
\left[ \frac{| K^S-\alpha|^2}{ K_{\bar{S}}^S} -3 \right]  \;\;. 
\label{dd}
\ee

To begin the analysis let us first define the form of these
corrections. We essentially follow the parametrization of 
ref.~\cite{casas96}, namely we shall study the following two cases:
\be
K({\rm Re}S) = K_0({\rm Re}S) + K_{np}({\rm Re}S)
\label{a1}
\ee
and
\be
e^{K({\rm Re}S)} = e^{K_0({\rm Re}S)} + e^{K_{np}({\rm Re}S)}
\label{a2}
\ee
with $K_0=-\log(2 {\rm Re}S)$ the usual tree level piece (we shall
neglect 1--loop corrections as they are irrelevant for the discussion)
and where $K_{np}$ in (\ref{a1}), and $e^{K_{np}}$ in (\ref{a2}) are 
given by \cite{shenk90}:
\be
 d g^{-p} e^{-b/g} = d ({\rm Re}S)^{p/2} e^{-b \sqrt{{\rm Re}S}}
\label{Knp}
\ee
where $d$, $p$ and $b$ are constants and we impose $b>0$ in order for
the correction to vanish in the weak coupling limit (i.e., when ${\rm
Re} S \rightarrow \infty$)\footnote{Notice that we do not impose any
requirement on the sign of $p$, that is on the behaviour of the 
non-perturbative correction at strong coupling.}.

Our goal is to perform a full analysis of the parameter space
determined by the three unknown constants, $d$, $p$ and $b$, in order
to find under which conditions the scalar potential will have a
minimum at a realistic value for the dilaton (${\rm Re} S \sim 2$ in
Planck units), with a reasonable gravitino mass ($m_{3/2} = M_P e^{K/2}
|W| \sim 1$ TeV) and, ideally, zero cosmological constant. 
For that purpose it is useful to realise that the presence of a second
term in the K\"ahler potential implies that for some values of these
parameters its second derivative may vanish. Notice that such a
derivative is the factor that multiplies the kinetic term for $S$ in
the SUGRA Lagrangian, and therefore the requirement of having a
physically meaningful dilaton imposes $K^{S}_{\bar{S}} >0$ (or, 
equivalently, $K''>0$, where $'$ denotes derivative with respect to
${\rm Re} S$). 
\begin{figure}
\centerline{  
\psfig{figure=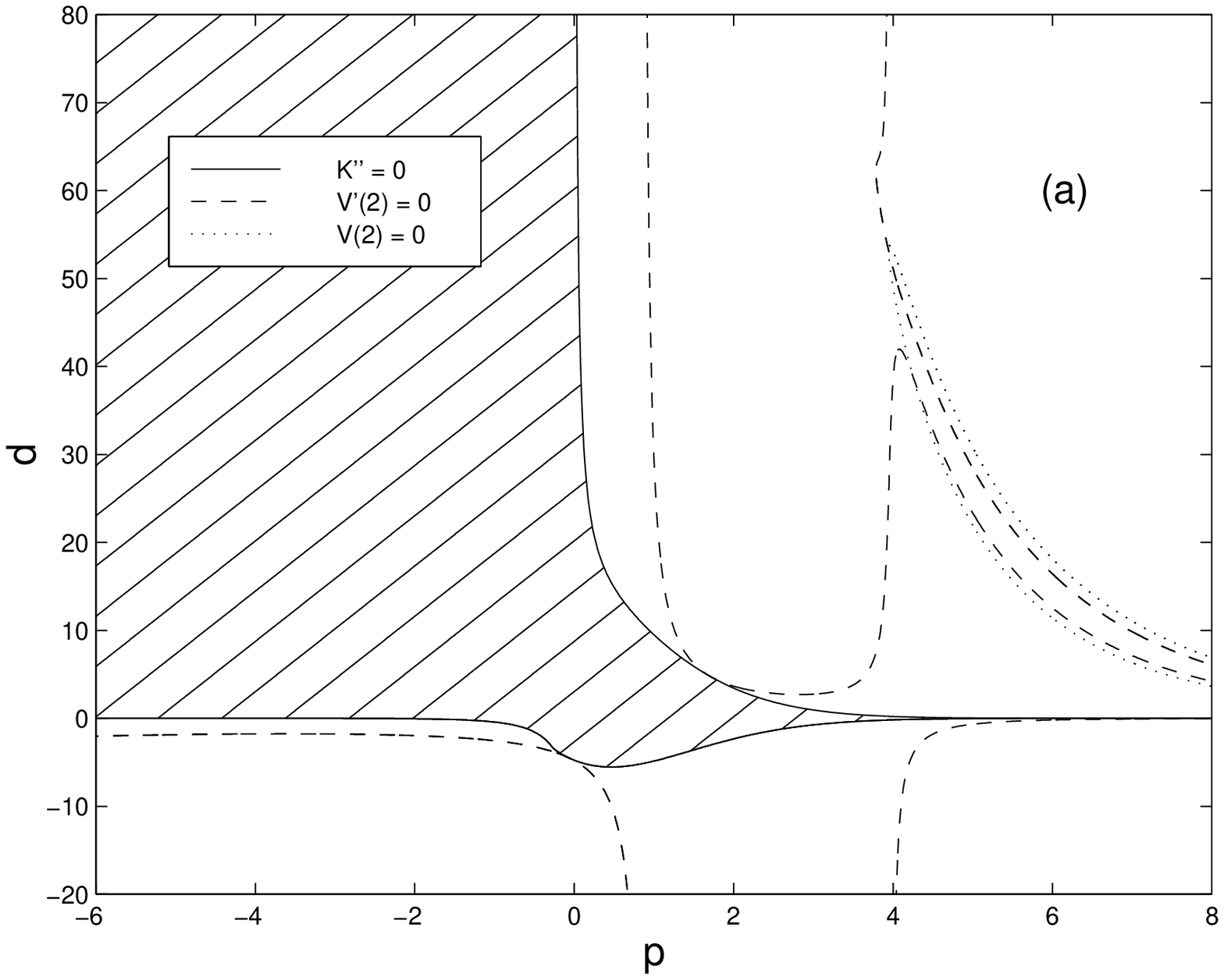,height=9cm,width=9cm,bbllx=0cm,bblly=7cm,bburx=21cm
,bbury=21cm}\
\psfig{figure=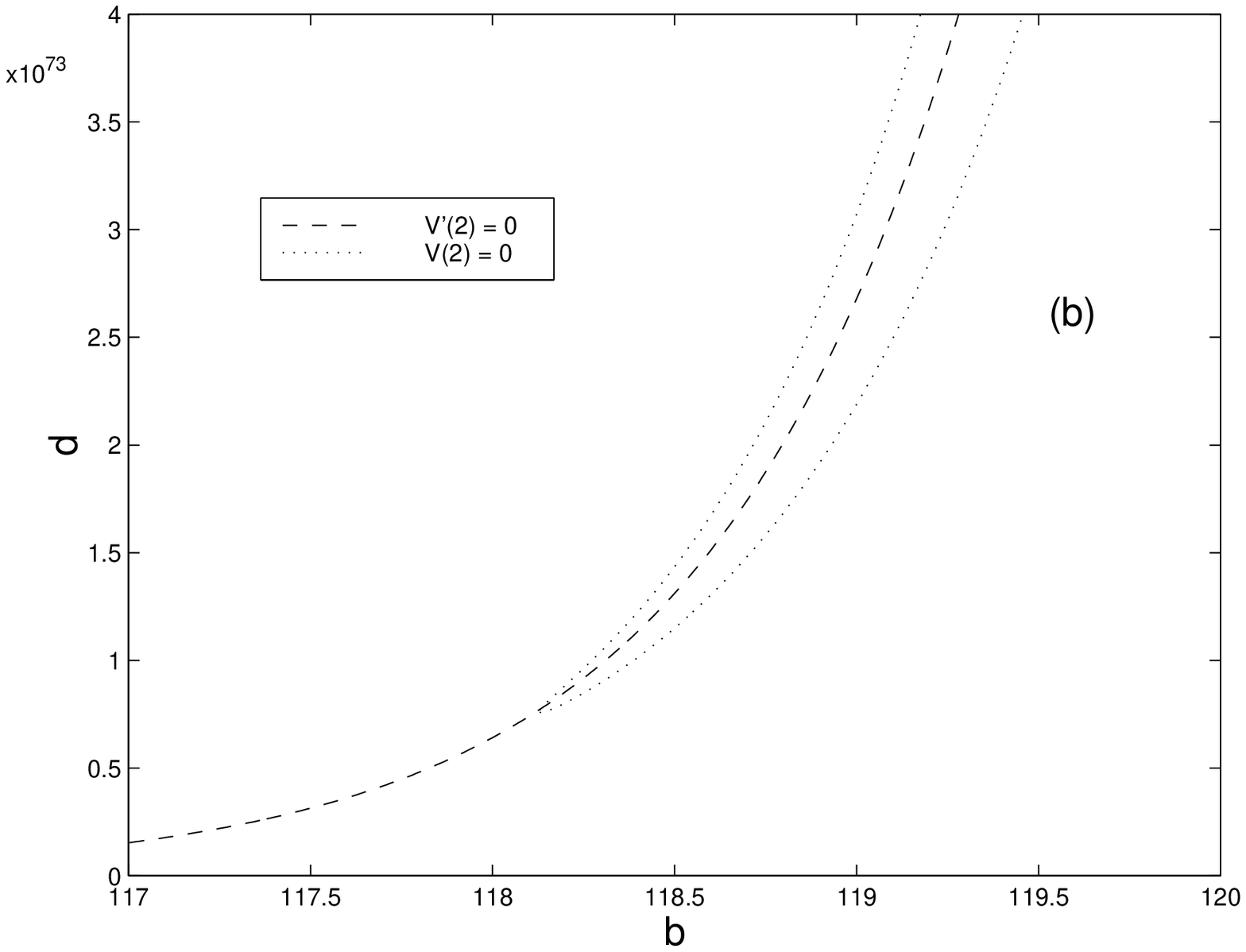,height=9cm,width=9cm,bbllx=0cm,bblly=7cm,bburx=21cm
,bbury=21cm}
}
\caption{}
{\footnotesize {\bf (a)} Contour lines of $K''=0$ (solid lines), 
$V'=0$ (dashed lines) and $V=0$ (dotted lines) in the $p$ vs 
$d$ plane for $b=1$, ${\rm Re} S \sim 2$ and gauge group SU(5). The 
hatched region corresponds to the requirement $K''>0$; {\bf (b)} Detail
of a region with $p=-3$ and $K''>0$ in the $b$ vs $d$ plane and gauge 
group SU(5). As before the dashed lines indicate contours of $V'=0$ 
and the dotted ones, contours of $V=0$.}
\end{figure}
This constraint is shown in figure~1~(a) for the case defined by 
eq.~(\ref{a1}), where the condition $K''=0$ for ${\rm Re} S =2$ 
represented by the solid lines determines the physically meaningful 
region (i.e. the hatched region) in the plane $p$ vs $d$, for
reasonable values of the latter; here we have
taken $b=1$ and a typical
gauge group, SU(5), which basically ensures a reasonable gravitino 
mass at the minimum of the potential. Unfortunately it can be seen 
that a simultaneous solution to the conditions $V'({\rm Re}S=2)=0$ 
(dashed lines) and $V({\rm Re}S=2)=0$ (dotted lines) always lies
within the unphysical region. At most we can have extrema which are 
compatible with the condition $K''=0$; these correspond to the
examples of minima next to a singularity of the potential already 
mentioned in ref.~\cite{casas96}. Moreover we have checked that it is 
possible to fine-tune the value of $d$ to a number of decimal figures 
so that the singularity reduces to a maximum, its height depending on 
the amount of fine-tuning we impose. However in all cases the value of
the cosmological constant is always positive and large.

There do exist solutions in which a minimum with zero cosmological 
constant occurs within the physical region, but they correspond
to very large values of the parameters $d$ and $p$ ($d \sim
10^{32}$, $p \sim -210$). We have checked that these minima with 
zero cosmological constant can be found for more reasonable values of 
the parameters by changing the gauge group. For example $d \sim 130$, 
$p \sim -10$ would correspond to a case with SU(80) ($b=1$) in the
hidden sector, which, unfortunately, implies a very big gravitino mass.

Fixing $p$ and changing $b$ leads to the same kind of picture emerging
as can be seen in fig.~1~(b), where we
have focussed on the physically allowed region which admits a minimum
with zero cosmological constant for $d \sim 7 \times 10^{72}$ and $b \sim 
118$. It corresponds to $p=-3$ and gauge group SU(5), and the 
corresponding gravitino mass is reasonable ($\sim 100$ GeV). It is 
convenient to keep in mind this 
particular example as we shall come back to it in the next section.
In general these solutions are characterised by extremely large values
of $d$, which make them look not too attractive. However we could
always rewrite the ansatz in a different way so that $d$ gets
redefined in terms of another constant which acquires then
reasonable values. 
\begin{figure}
\centerline{
\psfig{figure=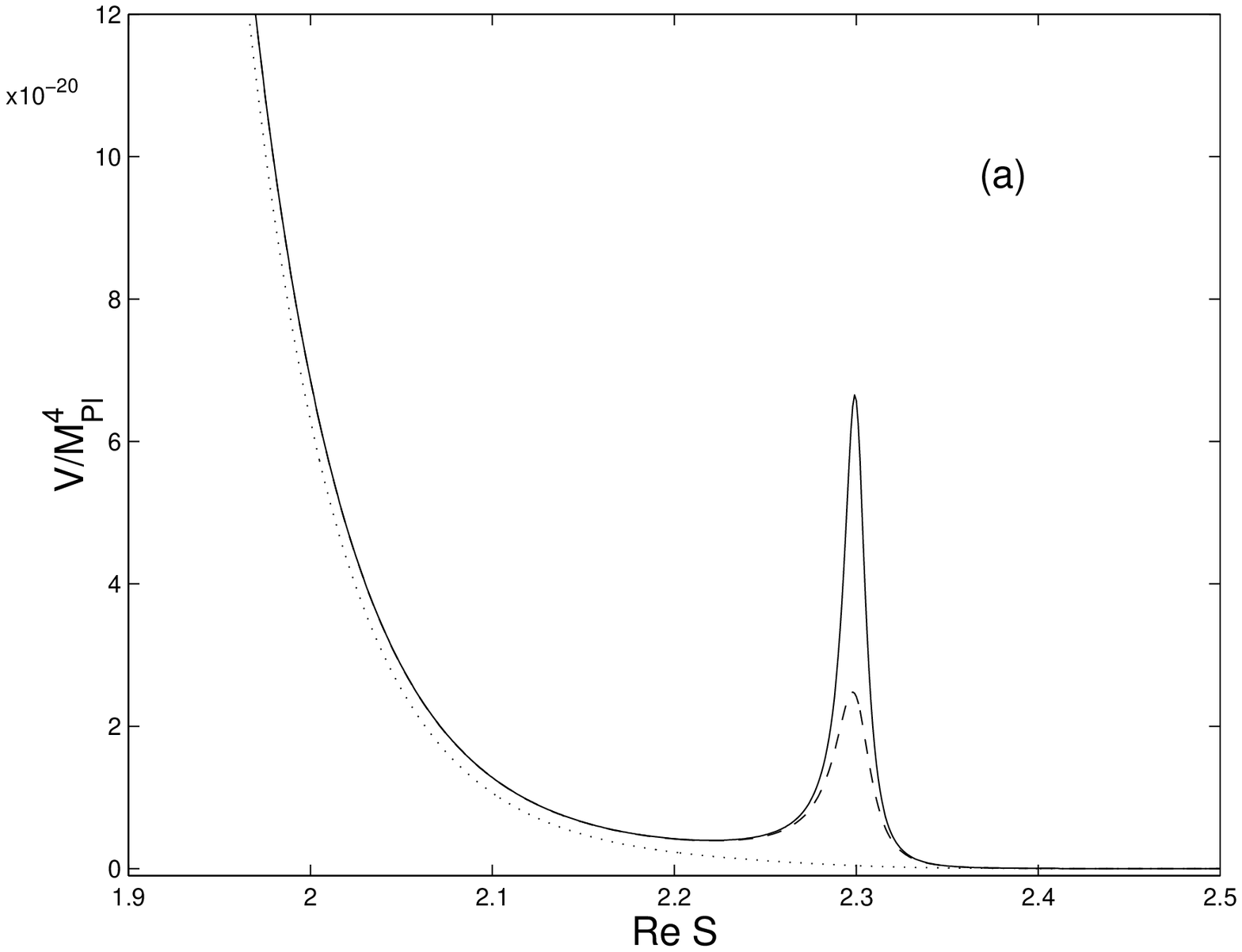,height=9cm,width=9cm,bbllx=0cm,bblly=7cm,bburx=21cm
,bbury=21cm}\
\psfig{figure=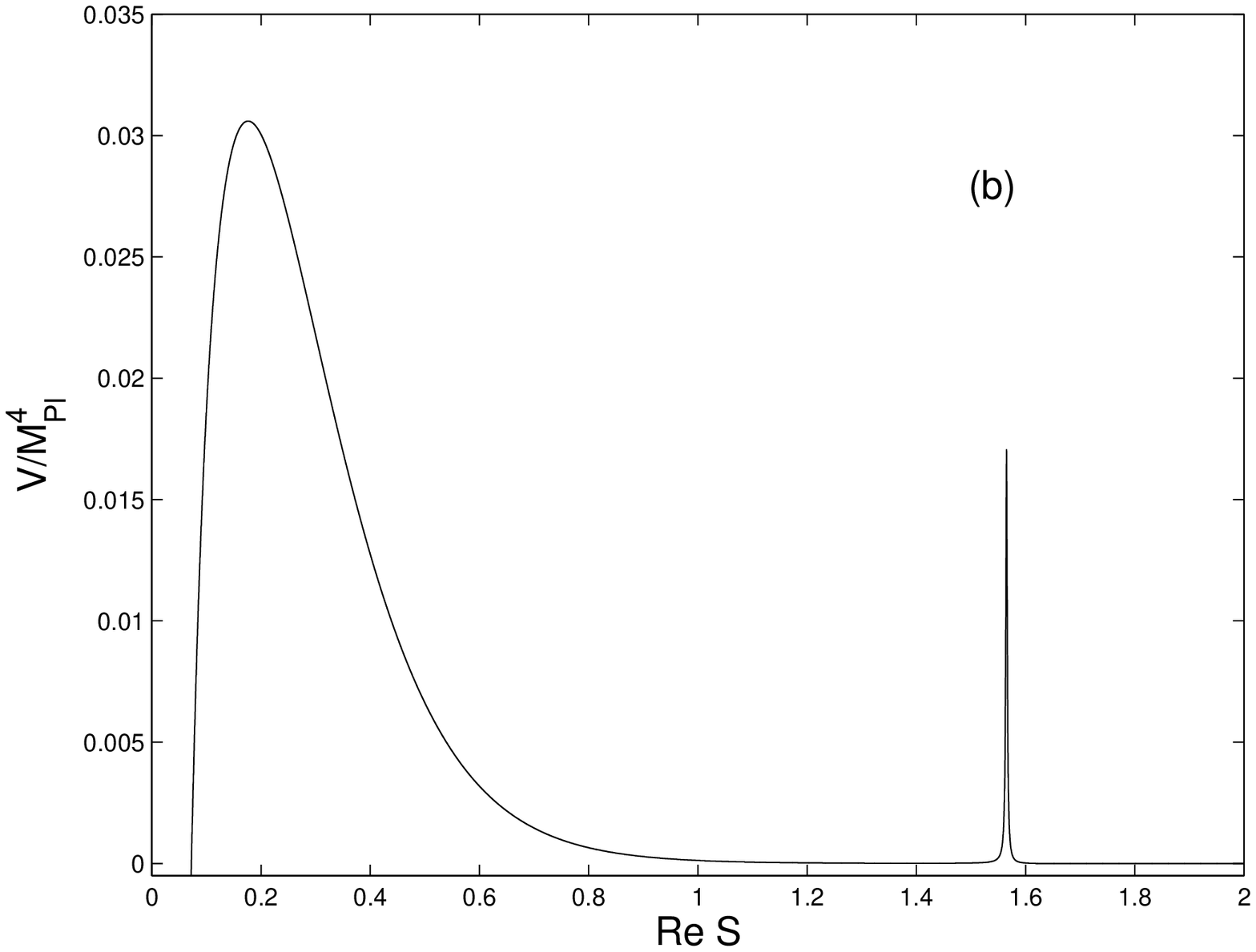,height=9cm,width=9cm,bbllx=0cm,bblly=7cm,bburx=21cm
,bbury=21cm}
}
\caption{}
{\footnotesize Plot of the scalar potential $V$ vs ${\rm Re}
S$ for one condensate and $K$ given by eq.~(\ref{a2}). The values of the
different parameters are: {\bf (a)} $p=1.1$, $b=1$, and $d=5.7391$ (solid),
$d=5.739$ (dashed), $d=5.73$ (dotted). The gauge group is SU(6). {\bf (b)} 
$p=1$, $b=1.1$, and $d=10.3775$. The gauge group is SU(11).}
\end{figure}

Turning to the second form for $K_{np}$, eq.~(\ref{a2}), the situation 
is much more complicated from the point of view of obtaining an
analytic solution to the conditions $V=V'=0$ for a reasonable value of
${\rm Re} S$ and $K''>0$. Therefore we are unable to produce plots
analogous to fig.~1. From eq.~(\ref{a2}) $e^{K}$ has to be positive 
definite, consequently there is always a lower bound on the possible 
values for $d$. As before we find that minima with positive $V_0$ are 
always associated to a singularity of the potential defined by
$K''=0$. Analogously to the case defined by eq.~(\ref{a1}), we
can fine-tune the value of the parameters to reduce the singularity to
a maximum, as shown in figure~2.
It is clear that, depending on the accuracy with
which we define $d$, the height of the barrier will change. This is
particularly relevant for the cosmological properties of these kind of
potentials, which we shall discuss below. Figure~2, case (a) corresponds to
a situation in which both the VEV of the dilaton and the gravitino 
mass are reasonable (i.e. $\sim 2.22$ and $\sim 5$ TeV,
respectively), whereas in case (b) the minimum corresponds to a 
phenomenologically acceptable value for the dilaton ($\sim 1.4$)
with the gravitino mass too large ($\sim 10^{12}$ GeV) to give rise to
a realistic spectrum at low energies. 

As mentioned in the introduction, we are interested in studying 
the shape of the dilaton potential in order to find out whether these 
non--perturbative corrections would alleviate its steepness, which is
one of the main problems pointed out in \cite{brust93} to build a 
realistic model of inflation based on superstrings. The answer to that
question is again given by fig.~2. There we can see that, as already
mentioned in \cite{brust93}, the bigger the gauge group is, the more
favourable the situation becomes in order to trap the dilaton in its 
minimum when it is rolling down from the strong coupling regime. As a
difference with \cite{brust93} the cosmological constant is positive
now and also the presence of the singularity allows us to tune the
height of the barrier to a certain extent. Case 2~(a) is analogous to
the several condensate case studied in \cite{brust93} where the
steepness of the potential for small values of $S$ prevents the
dilaton from settling down at the minimum. However, fig.~2~(b)
shows a nice shape for the potential, which would eventually drag the
dilaton to its minimum independently of the initial conditions, but   
at the price of not having a successful phenomenology. 
Alternatively it is known that exponential potentials allow the
existence of scaling solutions of the corresponding scalar field 
\cite{wette88}. We are currently exploring the possibility that such a
mechanism could stabilise the dilaton in the context of 
inflation \cite{barre98}.
  
However there is a second cosmological issue which might be solved in
the context of non-perturbative corrections and it concerns the 
so-called moduli problem \cite{decar93}. It is 
well-known that the masses of the dilaton and moduli in a scenario of 
SUSY breaking in the context of superstrings with zero cosmological
constant are of the order of the gravitino mass, and that causes
potential problems given their late decay; in general we can express
the mass of the dilaton as:
\be 
m_{S}^2 = \left. \frac{V''}{K''} \right|_{min}
= \frac{V_0}{M_P^2} \frac{P}{(K'')^3} + m_{3/2}^2 
\frac{Q}{(K'')^3} \;\;,
\label{dilmass}
\ee
where $V_0$ is the potential energy at the minimum, and $P$ and $Q$ 
are given by:
\be 
P = (K'')^3 - K''''K'' + (K''')^2 \;\;,
\ee
\be
Q = 2 (K'')^2 \left( (K' - 2 \alpha) + K'' \right) 
- 3 \left( K''' K'' (K' - 2 \alpha) + K'' K'''' - (K''')^2 \right) \;\;,
\ee
where $\alpha$ is defined after eq.~(\ref{sup}).
The crucial point here is to recall that the minimum of the potential
in our case lies very close to the point at which $K''=0$. This means
that the presence of the $(K'')^{-1}$ factor (which comes from the 
normalization of the kinetic term for the $S$ field) induces an 
enhancement in the mass of several orders of magnitude. For example, 
in case 2~(a) $m_{S} \sim 2 \times 10^{7} m_{3/2}$, with $1/K'' \sim
10^5$, whereas in 2~(b), $m_{S} \sim 2 \times 10^{5} m_{3/2}$, with 
$1/K'' \sim 5 \times 10^3$ (note that the contribution from the term 
proportional to the cosmological constant is the dominant one).
  
In conclusion, we have seen that it is possible to obtain solutions to
some of the most interesting problems associated to the phenomenology 
and cosmology of string-inspired models with these naive ansatze for
non-perturbative corrections to the K\"ahler potential. Therefore it
is reasonable to think that more elaborate forms for $K_{np}$ could
lead to even more promising results, and in this framework we proceed
to study other proposals.

\section{The linear multiplet formalism}

Having discussed in detail the ansatze for non--perturbative
corrections to the K\"ahler potential outlined in \cite{casas96},
let us turn to analyse other proposals made in the context of the
linear multiplet formalism \cite{binet97a,binet97b}. Following the results of 
Burgess et al.~\cite{burge95}, we know that both formulations, chiral
and linear, must be equivalent and our task is to show it explicitly
in this context. 

The K\"ahler potential in the linear multiplet formulation can be
expressed as
\be
K = \log(l) + g(l) \;\;,
\label{Kl}
\ee
where $l$ is the dilaton field, the lowest component of the vector
superfield which parametrises the gaugino condensate. At tree level,
and in the absence of non--perturbative corrections, $g(l)=0$ in the
expression above and, moreover, $1/l=2 {\rm Re} S$, recovering 
the usual tree--level K\"ahler potential in the chiral formulation. 
If we include non-perturbative corrections then, as indicated in
\cite{binet97b}, the relation between $l$ and $S$ changes to
\be
\frac{1+ f(l)}{l} = 2 {\rm Re} S
\label{lS}
\ee
where $f$ is related to $g$ via:
\be
\frac{dg(l)}{dl} = \frac{f(l)}{l} - \frac{df(l)}{dl}
\label{fg}
\ee
and, from these two equations, we can split eq.~(\ref{Kl}) in the same
form as eq.~(\ref{a1}), with the non-perturbative part given 
now by:
\be
K_{np} (l) = \log \left[ 1+f(l) \right] + g(l)  \;\;,
\label{npl}
\ee
which can be at least numerically expressed in terms of $S$ by using
eq.~(\ref{lS}).

The procedure in refs.~\cite{binet97a,binet97b}, once the formalism has been 
described, is to postulate an ansazt for $f(l)$, in particular we have
considered (i) $f(l) = A e^{-B/l}$ \cite{binet97a}, 
and (ii) $f(l) = (A_0+A_1/\sqrt{l}) e^{-B/\sqrt{l}}$
\cite{binet97b}. The first thing to note is that, in order to obtain
an expression for $K_{np}$, we must first integrate
eq.~(\ref{fg}) and, given (i) and (ii), the result is expressed
as an Exponential-Integral function, $E_i(x)$, which always admits an
infinite series expansion. 

We have reproduced exactly the results of refs.~\cite{binet97a,binet97b} 
for the
shape of the potential and position of the minima in both cases, (i)
and (ii). Moreover, we obtain identical results by numerically expressing
$l$ in terms of $S$, through eq.~(\ref{lS}), and using the {\em chiral} 
formulation to calculate the scalar potential and its minimum. Both
results totally agree proving, at least numerically, the equivalence 
between both formulations. It is also interesting to note that the
shape of these potentials is very much like the one shown in fig.~2,
including the scaling with the gauge group, with the difference that
now it is possible to tune the cosmological constant to zero. So it 
looks like two apparently different formulations and 
ansatze would lead to potentials with analogous characteristics.

\begin{figure}
\centerline{
\psfig{figure=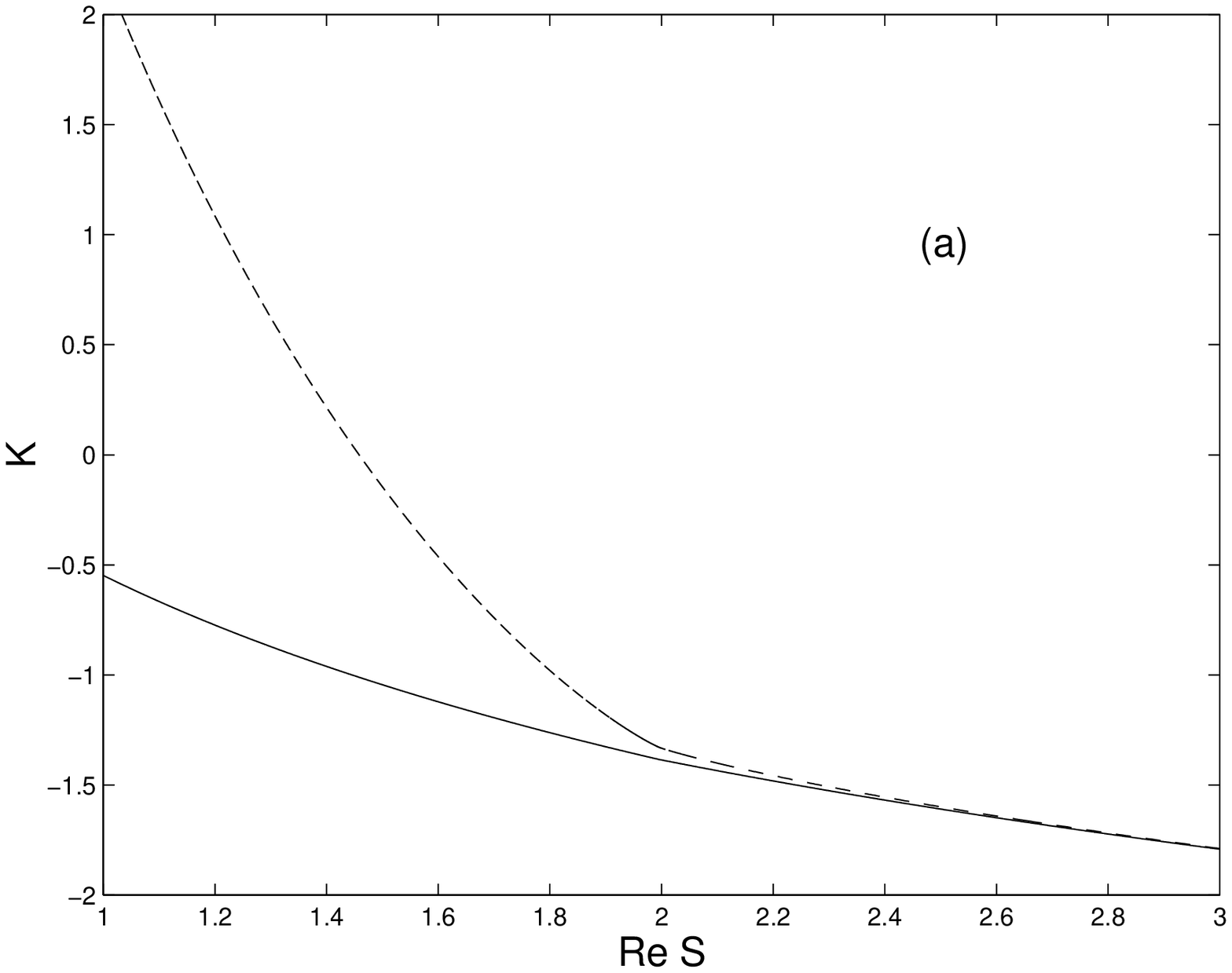,height=9cm,width=9cm,bbllx=0cm,bblly=7cm,bburx=21cm
,bbury=21cm}\
\psfig{figure=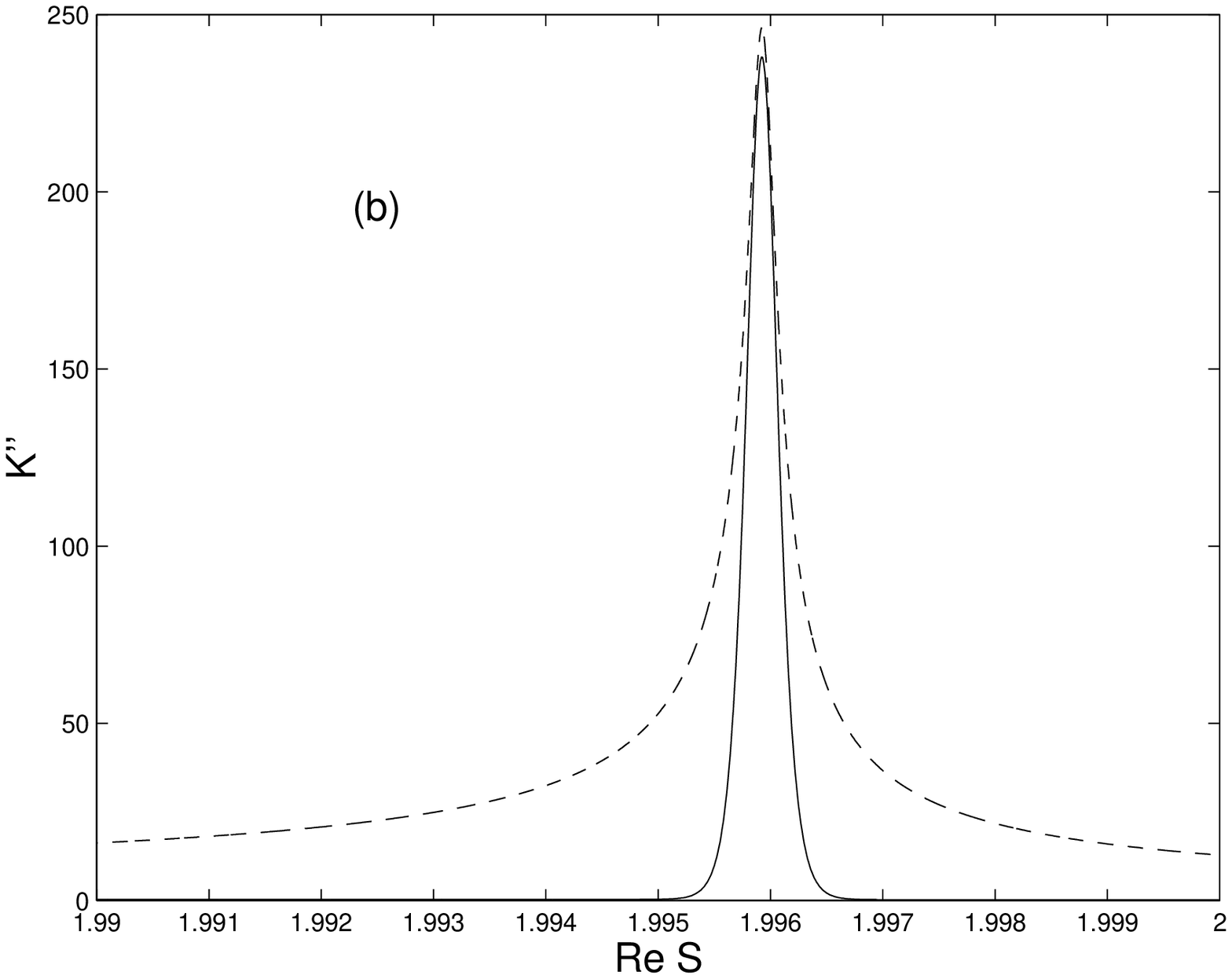,height=9cm,width=9cm,bbllx=0cm,bblly=7cm,bburx=21cm
,bbury=21cm}
}

\caption{}
{\footnotesize {\bf (a)} Plot of the K\"ahler potential $K$ vs 
${\rm Re} S$ for one condensate with gauge group $E_6$ and 2 matter 
multiplets. The dashed line represents the original ansatz (i) from 
ref~\cite{binet97a}, whereas the solid line is the result of using
our approximation eq.~(\ref{ours}) for $S_0 =2$ and $D=0.35$. $B \sim
30000$ is fixed by the condition of zero cosmological constant once 
$D$ has been specified; {\bf (b)} Same as in (a), but this 
time for $K_{S \bar{S}}$ vs ${\rm Re} S$.}
\end{figure}

Let's then analyse the form of $K_{np}$ in more detail. First of all,
eq.~(\ref{npl}) can be written in terms of $S$, and after some algebra
the result reduces to:
\be
K_{np} ({\rm Re}S) = \int_{{\rm Re} S}^{\infty} 
\frac{f({\rm Re}S')}{{\rm Re} S'} d({\rm Re} S') \;\;.
\label{npS}
\ee
Again we do not have an analytic expression for $f({\rm Re}S)$, but
even if we did, it still does not look clear what kind of relation,
if any, there could be between this and eq.~(\ref{Knp}).

Alternatively we can analyse the main features of the K\"{a}hler
potentials obtained in \cite{binet97a,binet97b}
and try to find an analytic functional form which may reproduce them.
For that purpose it is useful to realise that the examples
presented in \cite{binet97a,binet97b} share exactly the same 
properties, which
are depicted in fig.~3: the K\"ahler potential, shown in 3~(a) has
a ``kink'' at the point at which
the minimum develops, and its second derivative has a sharp maximum 
exactly at that point, as shown in 3~(b). This is exactly the opposite
behaviour to that of the examples in the previous section, where the 
minimum in the potential was due to $K''=0$ near $({\rm Re} S)_{min}$.
It also explains the mechanism to obtain a zero cosmological constant:
by adjusting the height of this maximum, the $1/K''$ term in 
eq.~(\ref{dd}) can have precisely the right magnitude to cancel 
the $-3|W|^2$ term.

These two very distinctive characteristics can be reproduced with the
following ansatz in the chiral formalism:
\be
K_{np} = \frac{D}{B\sqrt{{\rm Re} S}} 
\log\left( 1+e^{-B (\sqrt{{\rm Re} S}-\sqrt{S_0})} \right) 
\;\;,
\label{ours}
\ee
that depends on three parameters, $S_0$, $D$, and $B$, the first of
which just determines the value of ${\rm Re} S$ at the minimum.
Therefore this description is effectively made in terms of only $D$ 
and $B$, which are positive numbers. $D>0$ is imposed to avoid 
cancellations in $K''$, and therefore singularities in the scalar 
potential, and $B>0$ ensures the correct asymptotic behaviour in the 
weak coupling limit
(i.e. $K_{np} \rightarrow 0$ when ${\rm Re} S \rightarrow \infty$).
From figs.~3~(a),(b) we see that our ansatz eq.~(\ref{ours}), 
represented by 
the solid lines, reproduces the main features of the non-perturbative
corrections proposed in the linear formulation. In particular, we are
presenting here the results of the comparison with ansatz (i), but we 
have also reproduced the main features of (ii). Notice as well that
the numerical values of $D$ and $B$ are chosen such as to 
guarantee a zero cosmological constant, as shown in fig.~4~(a). In
order to compare with eq.~(\ref{Knp}) it is useful to split 
eq.~(\ref{ours}) into two different expressions, 
valid for ${\rm Re} S$ bigger/smaller than $S_0$. We find:
\begin{equation}
\begin{array}{cclr}
 e^{K_{np}} & = & 1 + \frac{D e^{B\sqrt{S_0}}}{B} ({\rm Re} S)^{-1/2} 
e^{-B \sqrt{{\rm Re} S}} \;,  &  {\rm Re} S > S_0 \;, \nonumber \\
\label{app} \\
 e^{K_{np}} & = & 
e^{-D (\sqrt{{\rm Re} S}-\sqrt{S_0})/\sqrt{{\rm Re} S}} \;, &
 {\rm Re} S < S_0 \;.\nonumber
\end{array}
\end{equation}
Using the previous expressions we can rewrite $e^K = e^{K_0+K_{np}}$
as $e^K = e^{K_0}+ e^{{\tilde{K}}_{np}}$, where now 
\begin{equation}
\begin{array}{cclr}
e^{{\tilde{K}}_{np}} & = & \frac{D e^{B\sqrt{S_0}}}{2B} ({\rm Re} S)^{-3/2} 
e^{-B \sqrt{{\rm Re} S}} \;, & {\rm Re} S > S_0 \;, \nonumber \\
\label{app2} \\
e^{{\tilde{K}}_{np}} & = & \frac{1}{2 {\rm Re} S} \left[ -1+
e^{-D (\sqrt{{\rm Re} S}-\sqrt{S_0})/\sqrt{{\rm Re} S}} \right]
 \sim  \frac{e^{-D}}{2} ({\rm Re} S)^{-1} 
e^{D \sqrt{S_0}/\sqrt{{\rm Re} S}} \;, & {\rm Re} S < S_0 \;,
\nonumber
\end{array}
\end{equation}
making explicit the resemblance of these expressions with that
of eq.~(\ref{Knp}) in the context of eq.~(\ref{a2}). Note that, in the
weak coupling regime (i.e. ${\rm Re} S > S_0$), we can make the
identifications $d = D e^{B\sqrt{S_0}}/2B$, $p=-3$ and $B=b$. In
particular the size of $B$ required to reproduce the linear multiplet
results makes $d$ extremely large, and that corresponds precisely to the
solution we found in the previous section with a realistic minimum 
and zero cosmological constant (i.e. fig.~1~(b)). Therefore we can 
conclude that the
proposals for non-perturbative corrections to the K\"ahler potential
made in the context of the {\em chiral} \cite{casas96} and {\em linear}
\cite{binet97a,binet97b} multiplet formalisms essentially correspond to a same 
functional form evaluated in different regions of parameter space.

One final issue we should address in the context of the linear
multiplet formalism is that of the dilaton mass. We saw in the
previous section that the vanishing of $K''$ around the minimum
implies an enhancement of the dilaton mass with respect to the
gravitino one due to the contribution of factors of $1/K''$ to the 
former. Now we are dealing with a different situation as $K''$ has a 
maximum at the minimum of the potential, therefore we would naively expect
some suppression of $m_S$ with respect to $m_{3/2}$.
\begin{figure}
\centerline{
\psfig{figure=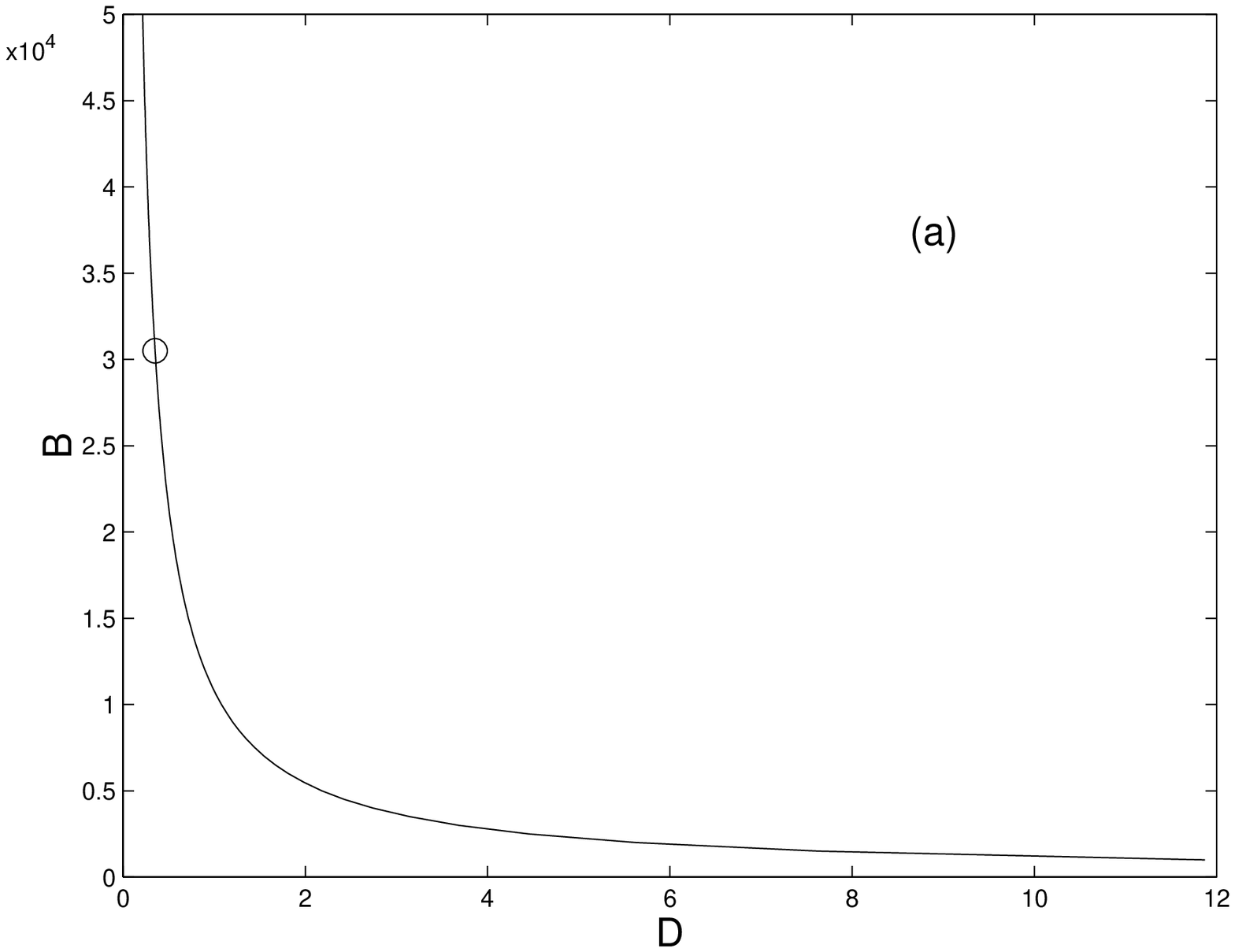,height=9cm,width=9cm,bbllx=0cm,bblly=7cm,bburx=21cm
,bbury=21cm}\
\psfig{figure=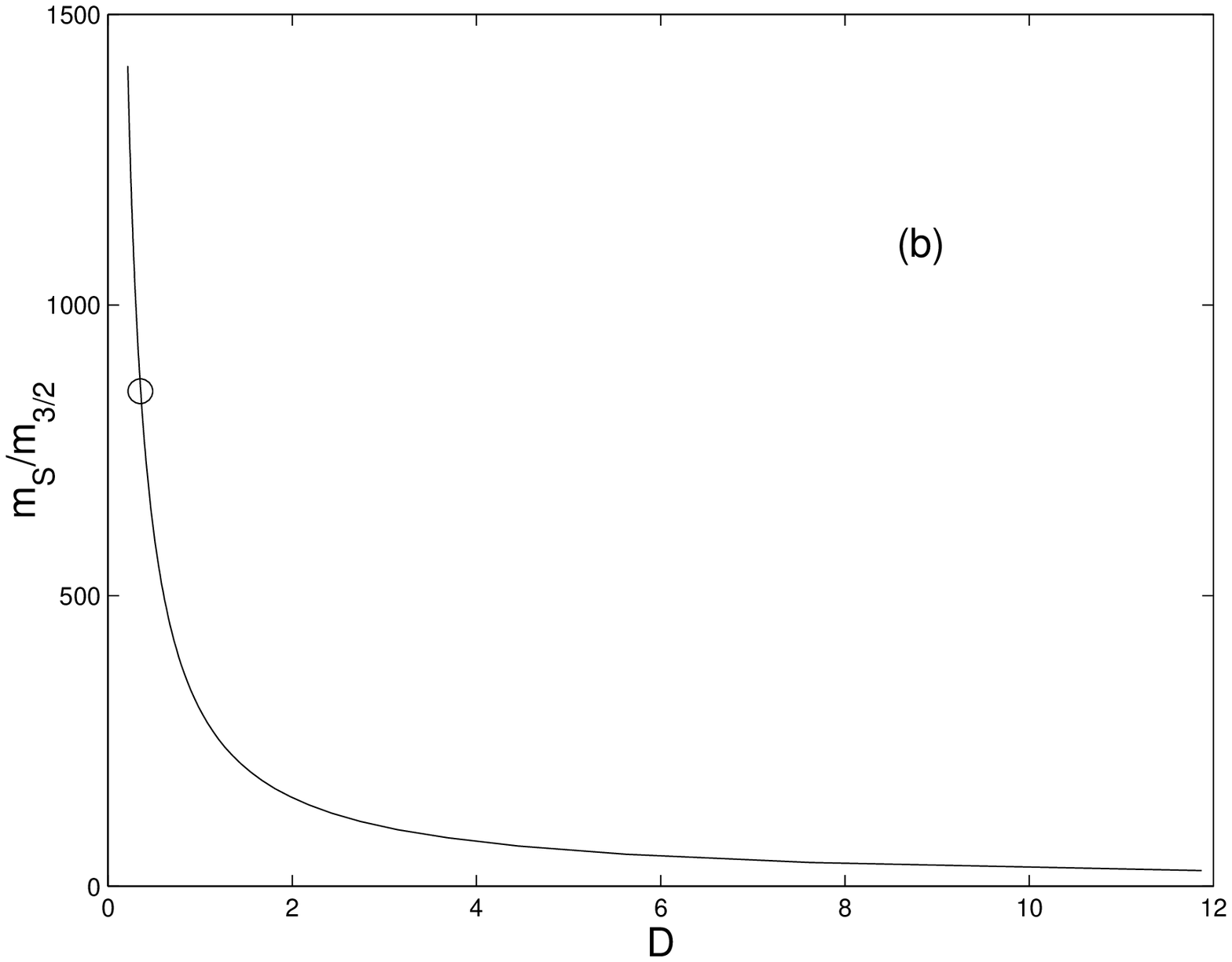,height=9cm,width=9cm,bbllx=0cm,bblly=7cm,bburx=21cm
,bbury=21cm}
}
\caption{}
{\footnotesize {\bf (a)} Values of the B and D parameters (solid line) that 
give a zero cosmological constant with our ansatz eq.~(\ref{ours}). 
The circle  
corresponds to the numbers we use to reproduce the results in 
ref.~\cite{binet97a};
{\bf (b)} plot of the ratio between the dilaton and
gravitino masses (solid line) versus the parameter $D$ using our
ansatz eq.~(\ref{ours}), with $S_0=2$ and $B$ fixed to give zero
cosmological constant. 
The circle corresponds to the result of
ref.~\cite{binet97a}.}
\end{figure}
However this is
not true in general, as shown in fig.~4~(b), where we plot the
ratio between the dilaton and gravitino masses, $(m_S/m_{3/2})$,
versus $D$, with the corresponding $B$ given by fig.~4~(a).
Therefore, in eq.~(\ref{dilmass}) only the second term contributes,
and the coefficient $Q$ is totally dominated by the fourth derivative
of the K\"ahler potential. So the dilaton mass in these models is
given by:
\be
m_{S}^2 = m_{3/2}^2 \frac{-3 K''''}{(K'')^2}
\ee
As we can see from fig.~4~(b), the closer $D$ is to zero, the bigger the
enhancement factor of $m_S$ with respect to $m_{3/2}$ is, and it can
amount to about three orders of magnitude. The circle represents
the result when we use the original ansatz of ref.~\cite{binet97a}.
We can obtain with our ansatz bigger values than that, but it implies
extremely large values for $B$, as can be seen in fig~4~(a).

\section{Conclusions}

We have performed an exhaustive study of the effects of string
motivated non-perturbative corrections to the K\"ahler potential
on the phenomenology of superstring derived models. Using previous
proposals made in the context of the chiral formulation of gaugino
condensation \cite{casas96} (with the dilaton as part of a chiral 
supermultiplet),
we have fully analyzed the parameter space which they define, and
we have shown that a minimum of the scalar potential with a reasonable
value for the dilaton and zero cosmological constant can only be
achieved for extremely unreasonable values of at least one of the 
parameters. On the other hand, reasonable values for these give
aceptable minima but with positive cosmological constant. In any
case, the shape of these potentials does not look very promising as
their steepness in the strong coupling regime would prevent the
dilaton from settling down at its minimum. However we have shown that 
the so-called
moduli problem gets alleviated, at least in the dilaton sector.

Once we had fully explored these simpler proposals we turned to study 
the results obtained in the context of
the linear multiplet formalism to describe gaugino condensation.
We have shown that the ansazte proposed so far in that context can 
be at least numerically reproduced using the chiral formalism and,
moreover, we have proposed an explicit ansatz which essentially 
reproduces in chiral language the main features of these 
non-peturbative corrections proposed in \cite{binet97a,binet97b}.
To a high degree of accuracy, this ansatz 
looks very similar to the one initially studied in the chiral
formulation, and the corresponding values of the parameters that
define them seem to indicate that we are now describing the same
type of solutions previously found with zero cosmological constant. 
We have shown that both formulations lead to essentially identical 
results and, in particular, that the results obtained by \cite{casas96} and 
\cite{binet97a,binet97b} correspond to basically the same ansatz 
evaluated in different regions of parameter space. Moreover we are 
able to reproduce a large hierarchy between the dilaton and gravitino 
masses, opening up a promising way of solving the usual cosmological 
problems associated with string-inspired models.

\section*{Acknowledgements}

We thank David Bailin for very interesting discussions. B.deC. thanks 
Alberto Casas and Fernando Quevedo for their encouragement and 
invaluable advice.

\end{document}